\documentclass{article}

\usepackage{graphicx}
\usepackage{vmargin}
\usepackage{cite}
\setmarginsrb{2.5cm}{1.5cm}{2.cm}{2.25cm}{1cm}{1cm}{1cm}{1.5cm}

\begin{document}

\title{{\bf Deeply Virtual Compton Scattering and Meson Production at JLab/CLAS}}


\author{Hyon-Suk Jo\\
\\
Institut de Physique Nucl\'{e}aire d'Orsay, 91406 Orsay, France}

\date{}

\maketitle

\begin{abstract}
This report reviews the recent experimental results from the CLAS
collaboration (Hall B of Jefferson Lab, or JLab) on Deeply Virtual
Compton Scattering (DVCS) and Deeply Virtual Meson Production (DVMP)
and discusses their interpretation in the framework of Generalized
Parton Distributions (GPDs). The impact of the experimental data on
the applicability of the GPD mechanism to these exclusive reactions
is discussed. Initial results obtained from JLab 6~GeV data indicate
that DVCS might already be interpretable in this framework while
GPD models fail to describe the exclusive meson production (DVMP)
data with the GPD parameterizations presently used. An exception is
the $\phi$ meson production for which the GPD mechanism appears
to apply. The recent global analyses aiming to extract GPDs from
fitting DVCS CLAS and world data are discussed. The GPD experimental
program at CLAS12, planned with the upcoming 12~GeV upgrade of
JLab, is briefly presented.
\end{abstract}

\section*{Introduction}

Generalized Parton Distributions take the description of the
complex internal structure of the nucleon to a new level by providing
access to, among other things, the correlations between the (transverse)
position and (longitudinal) momentum distributions of the partons in
the nucleon. They also give access to the orbital momentum contribution
of partons to the spin of the nucleon. 

GPDs can be accessed via Deeply Virtual Compton Scattering and exclusive
meson electroproduction, processes where an electron interacts with a
parton from the nucleon by the exchange of a virtual photon and that
parton radiates a real photon (in the case of DVCS) or hadronizes into
a meson (in the case of DVMP). The amplitude of the studied process can
be factorized into a hard-scattering part, exactly calculable in pQCD
or QED, and a non-perturbative part, representing the soft structure of
the nucleon, parametrized by the GPDs. At leading twist and leading order
approximation, there are four independent quark helicity conserving GPDs
for the nucleon: $H$, $E$, $\tilde{H}$ and $\tilde{E}$. These GPDs are
functions depending on three variables $x$, $\xi$ and $t$, among which
only $\xi$ and $t$ are experimentally accessible. The quantities $x+\xi$
and $x-\xi$ represent respectively the longitudinal momentum fractions
carried by the initial and final parton. The variable $\xi$ is linked
to the Bjorken variable $x_{B}$ through the asymptotic formula:
$\xi=\frac{x_{B}}{2-x_{B}}$. The variable $t$ is the squared momentum
transfer between the initial and final nucleon. Since the variable $x$
is not experimentally accessible, only Compton Form Factors, or CFFs
(${\cal H}$, ${\cal E}$, $\tilde{{\cal H}}$ and $\tilde{{\cal E}}$),
which real parts are weighted integrals of GPDs over $x$ and imaginary
parts are combinations of GPDs at the lines $x=\pm\xi$, can be extracted.

The reader is referred to Refs. \cite{gpd1, gpd2, gpd3, gpd4, gpd5,
gpd6, gpd7, gpd8, vgg1, vgg2, bmk} for detailed reviews on the
GPDs and the theoretical formalism.

\section*{Deeply Virtual Compton Scattering}

\begin{figure}[t]
  \centerline{\includegraphics[height=0.11\textheight]{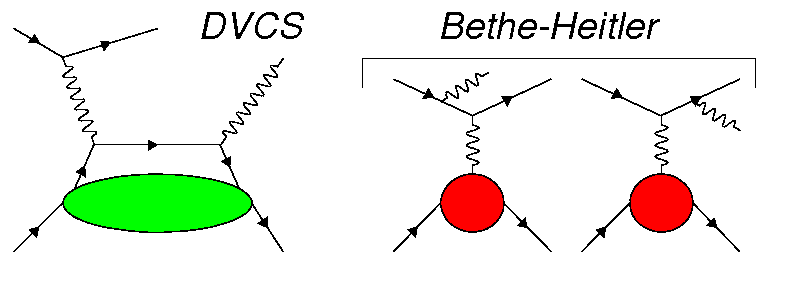}}
  \caption{Handbag diagram for DVCS (left) and diagrams for
  Bethe-Heitler (right), the two processes contributing to the
  amplitude of the $eN \to eN\gamma$ reaction.}
  \label{fig:diagrams}
\end{figure}

Among the exclusive reactions allowing access to GPDs, Deeply Virtual
Compton Scattering (DVCS), which corresponds to the electroproduction
of a real photon off a nucleon $eN \to eN\gamma$, is the key reaction
since it offers the simplest, most straighforward theoretical
interpretation in terms of GPDs. The DVCS amplitude interferes with the
amplitude of the Bethe-Heitler (BH) process which leads to the exact same
final state. In the BH process, the real photon is emitted by either the
incoming or the scattered electron while in the case of DVCS, it is
emitted by the target nucleon (see Figure~\ref{fig:diagrams}). Although
these two processes are experimentally indistinguishable, the BH is well
known and exactly calculable in QED. At current JLab energies (6~GeV),
the BH process is highly dominant (in most of the phase space) but the
DVCS process can be accessed via the interference term rising from the
two processes. With a polarized beam or/and a polarized target, different
types of asymmetries can be extracted: beam-spin asymmetries ($A_{LU}$),
longitudinally polarized target-spin asymmetries ($A_{UL}$), transversely
polarized target-spin asymmetries ($A_{UT}$), double-spin asymmetries
($A_{LL}$, $A_{LT}$). Each type of asymmetry gives access to a different
combination of Compton Form Factors.

\vspace{0.5cm}
\begin{figure}[h]
  \centerline{\includegraphics[height=0.34\textheight]{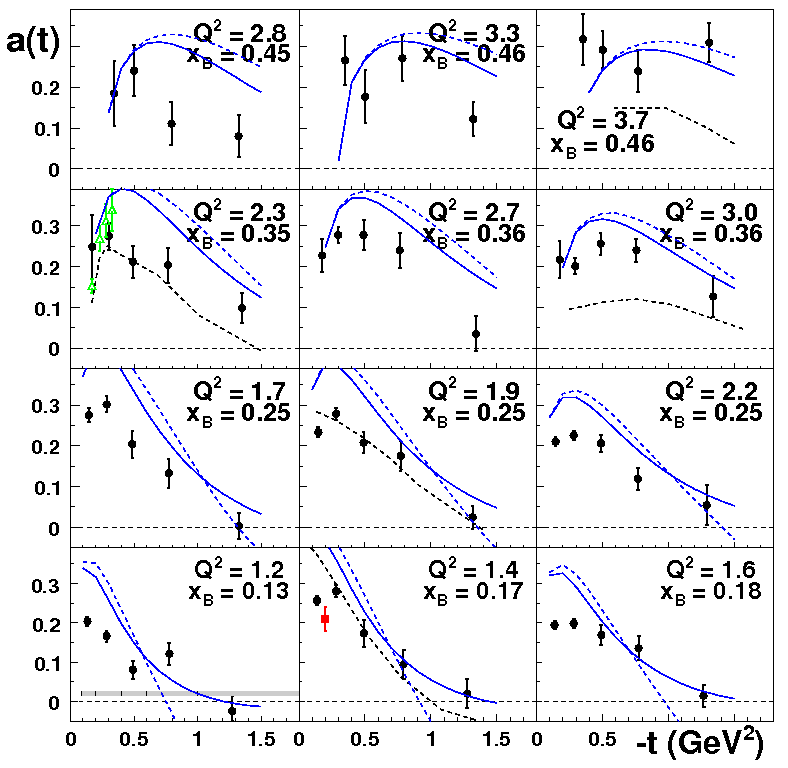}}
  \caption{DVCS beam-spin asymmetries as a function of $-t$, for
  different values of $Q^{2}$ and $x_{B}$. The (black) circles represent
  the latest CLAS results \cite{bsa3}, the (red) squares and the (green)
  triangles are the results, respectively, from Ref.~\cite{bsa1} and
  Ref.~\cite{halla}. The black dashed curves represent Regge calculations
  \cite{jml}. The blue curves correspond to the GPD calculations of
  Ref.~\cite{vgg1} (VGG) at twist-2 (solid) and twist-3 (dashed) levels,
  with the contribution of the GPD $H$ only.}
  \label{fig:bsa}
\end{figure}

\vspace{0.5cm}
The first results on DVCS beam-spin asymmetries published by the CLAS
collaboration were extracted using data from non-dedicated experiments
\cite{bsa1, bsa2}. Also using non-dedicated data, CLAS published DVCS
longitudinally polarized target-spin asymmetries in 2006 \cite{tsa}.
In 2005, the first part of the e1-DVCS experiment was carried out in
the Hall B of JLab using the CLAS spectrometer \cite{clas} and an
additional electromagnetic calorimeter, made of 424 lead-tungstate
scintillating crystals read out via avalanche photodiodes, specially
designed and built for the experiment. This additional calorimeter
was located at the forward angles, where the DVCS/BH photons are mostly
emitted, as the standard CLAS configuration does not allow detection
at those forward angles. This first CLAS experiment dedicated to DVCS
measurements, with this upgraded setup allowing a fully exclusive
measurement, ran using a 5.766~GeV polarized electron beam and a
liquid-hydrogen target. From this experiment data, CLAS published in
2008 the largest set of DVCS beam-spin asymmetries ever extracted in
the valence quark region \cite{bsa3}. Figure~\ref{fig:bsa} shows the
corresponding results as a function of $-t$ for different bins in
($Q^{2}$, $x_{B}$). The predictions using the GPD model from VGG
(Vanderhaeghen, Guichon, Guidal) \cite{vgg1, vgg2} overestimate
the asymmetries at low $|-t|$, especially for small values of $Q^{2}$
which can be expected since the GPD mechanism is supposed to be valid
at high $Q^{2}$. Regge calculations \cite{jml} are in fair agreement
with the results at low $Q^{2}$ but fail to describe them at high
$Q^{2}$ as expected. We are currently working on extracting DVCS
unpolarized and polarized absolute cross sections from the e1-DVCS
data \cite{hsj}.

Having both the beam-spin asymmetries and the longitudinally polarized
target-spin asymmetries, a largely model-independent GPD analysis in
leading twist was performed, fitting simultaneously the values for
$A_{LU}$ and $A_{UL}$ obtained with CLAS at three values of $t$ and fixed
$x_{B}$, to extract numerical constraints on the imaginary parts of the
Compton Form Factors (CFFs) ${\cal H}$ and $\tilde{{\cal H}}$, with
average uncertainties of the order of 30\% \cite{guidal_clas}. Before
that, the same analysis was performed fitting the DVCS unpolarized and
polarized cross sections published by the JLab Hall A collaboration
\cite{halla} to extract numerical constraints on the real and imaginary
parts of the CFF ${\cal H}$ \cite{guidal_fitter_code}. Another GPD
analysis in leading twist, assuming the dominance of the GPD $H$ (the
contributions of $\tilde{H}$, $E$ and $\tilde{E}$ being neglicted) and
using the CLAS $A_{LU}$ data as well as the DVCS JLab Hall A data, was
performed to extract constraints on the real and imaginary parts of
the CFF ${\cal H}$ \cite{moutarde}. Similar analyses were performed
using results published by the HERMES collaboration
\cite{guidal_moutarde, guidal_hermes}. A third approach was developped,
using a model-based global fit on the available world data to calculate
the real and imaginary parts of the CFF ${\cal H}$ \cite{km}.
When we compare the different results of those analyses for the
imaginary part of ${\cal H}$, they appear to be relatively compatible
(such a comparison plot can be found in Ref.~\cite{km2}).

\section*{Deeply Virtual Meson Production}

The CLAS collaboration published several results on pseudoscalar meson
electroproduction ($\pi^{0}$, $\pi^{+}$) \cite{ps1, ps2}. However, those
are not reviewed in this paper, limiting itself to vector mesons.

CLAS published cross-section measurements for the following vector mesons:
$\rho^{0}$ \cite{rho01, rho02}, $\omega$ \cite{omega} and $\phi$
\cite{phi1, phi2}, contributing significantly to the world data on
vector mesons with measurements in the valence quark region, corresponding
to low $W$ ($W<5$~GeV). First measurements of $\rho^{+}$ electroproduction
are being extracted from the e1-DVCS data mentionned above \cite{rho+}.

\begin{figure}[h]
  \centerline{\includegraphics[height=0.37\textheight]{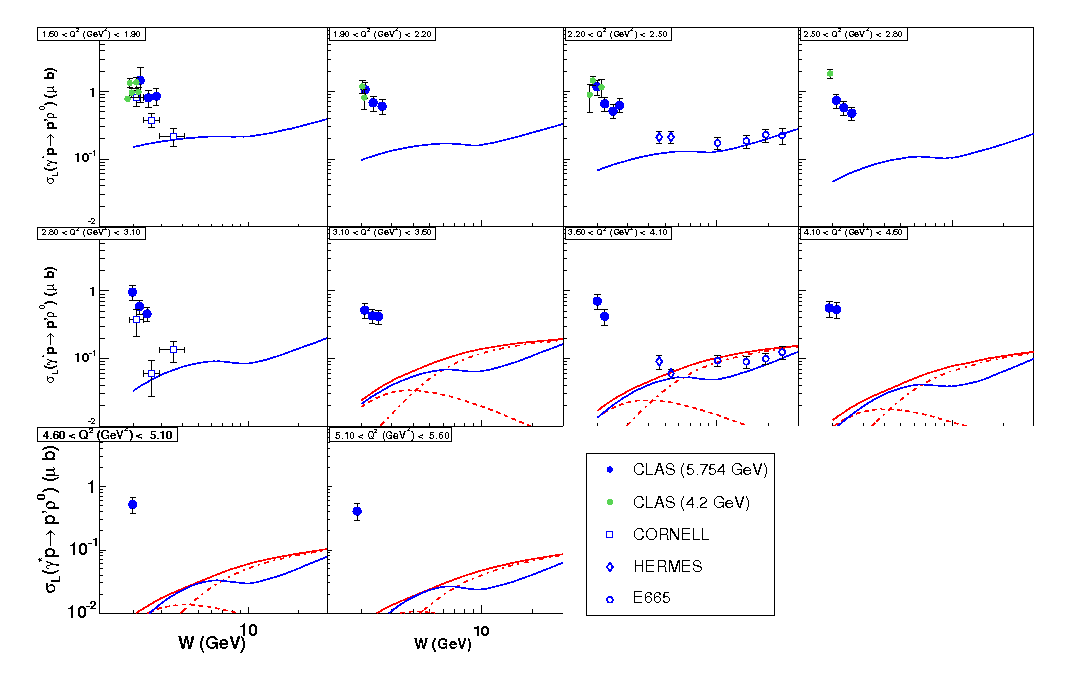}}
  \caption{Longitudinal cross sections for $\rho^{0}$ as a function of
  $W$ at fixed $Q^{2}$ (CLAS and world data). The results from CLAS are
  shown as full circles. The blue curves are VGG GPD-based predictions.
  The red curves represent GK GPD-based predictions: total (solid),
  valence quarks (dashed), sea quarks and gluons (dot-dashed).}
  \label{fig:rho}
\end{figure}

As the leading-twist handbag diagram is only valid for the longitudinal
part of the cross section of those vector mesons, it is required to
separate the longitudinal and transverse parts of the cross sections
extracted from the experimental data by analyzing the decay angular
distribution of the meson. Figure~\ref{fig:rho} shows the longitudinal
cross sections of the $\rho^{0}$ meson production
$\sigma_{L}(\gamma^{*} p \to p \rho^{0})$  as a function of $W$ at
fixed $Q^{2}$, for different bins in $Q^{2}$. As a function of
increasing $W$, those cross sections first drops at low $W$ ($W<5$~GeV,
corresponding to the valence quark region) and then slightly rise at
higher $W$. The longitudinal cross sections of the $\omega$ meson
production seems to show the same behavior as a function of $W$ as the
one observed for the $\rho^{0}$ meson. 
The GPD-based predictions from VGG and from GK (Goloskokov, Kroll)
\cite{gk} describe quite well those results at high $W$ but both
GPD models fail by large to reproduce the behavior at low $W$ (see the
curves on Figure~\ref{fig:rho}). The $\phi$ meson production, which is
mostly sensitive to gluon GPDs, is a different case as its longitudinal
cross sections as a function of $W$ show a different behavior by
continuously rising with increasing $W$ all the way from the lowest $W$
region; these cross sections are very well described by the GPD model
predictions \cite{gk_phi}. The reason why the GPD models fail to
describe the data for the $\rho^{0}$ and $\omega$ mesons at low $W$
(valence quark region) is unsure at this point. The handbag mechanism
might not be dominant in the low $W$ valence region as the minimum
value of $|-t|$ increases with decreasing $W$ and higher-twist effects
grow with $t$. Another possibility is that the handbag mechanism might
actually be dominant in the low $W$ valence region but there is an
important contribution missing in the GPD models.

\section*{DVCS and DVMP at CLAS12}

With the upcoming 12~GeV upgrade of JLab's CEBAF accelerator, the
instrumentation in the experimental halls will be upgraded as well. In
Hall B, the CLAS detector will be replaced by the new CLAS12 spectrometer,
under construction, with the study of Generalized Parton Distributions as
one of the highest priorities of its future experimental program. The
experiments currently proposed have the following goals:
\begin{itemize}
\item DVCS beam-spin asymmetries on the proton,
\item DVCS longitudinal target-spin asymmetries on the proton,
\item DVCS transverse target-spin asymmetries on the proton,
\item DVCS on the neutron,
\item DVCS unpolarized and polarized cross sections,
\item DVMP: pseudoscalar meson electroproduction,
\item DVMP: vector meson electroproduction.
\end{itemize}
To study DVCS on the neutron, a central neutron detector was designed to
be added to the base equipment of the CLAS12 spectrometer. A combined
analysis of DVCS on the proton and on the neutron allows flavor
separation of GPDs.

\vspace{0.3cm}
\begin{figure}[h]
  \centerline{\includegraphics[height=0.3\textheight]{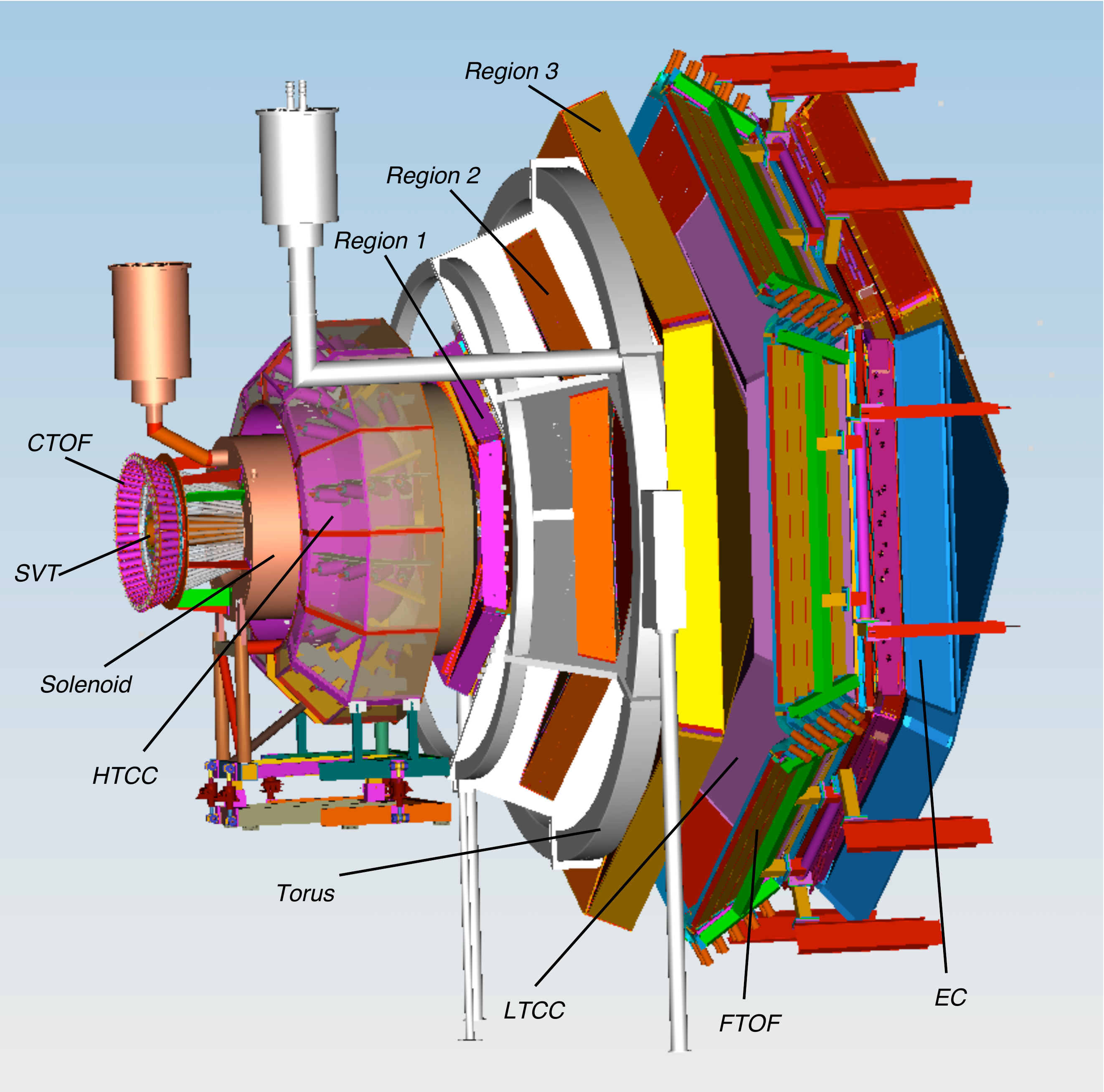}}
  \caption{The CLAS12 detector currently under construction.}
  \label{fig:clas12}
\end{figure}

JLab 12~GeV will provide high luminosity (L$\sim10^{35}$cm$^{-2}$s$^{-1}$)
for high accuracy measurements to study GPDs in the valence quark region
and test the models on a large $x_{B}$ scale. The new CLAS12 spectrometer,
with its large acceptance allowing measurements on a large kinematic range,
will be perfectly fitted for a rich GPD experimental program.

\section*{Conclusions}

The CLAS collaboration produced the largest set of data for DVCS and
exclusive vector meson production ever extracted in the valence quark region.
The VGG GPD model fairly agrees with the DVCS asymmetry data at high $Q^{2}$
but fails to reproduce it at lower $Q^{2}$. As for the exclusive vector meson
data, GPD models describe well the longitudinal cross sections at high $W$
(region corresponding to sea quarks and/or gluons) which seem to be
interpretable in terms of leading-twist handbag diagram (quark/gluon GPDs)
but fail by large for $W<5$~GeV (corresponding to the valence quark region)
except for the $\phi$ meson for which the GPD formalism seems to apply. We
need experimental data of higher $Q^{2}$ while staying in the valence quark
region to extend the DVCS data on a larger kinematic domain and provide more
constraints for the GPD models, and to test the GPD mechanism validity regime
for DVMP. JLab 12~GeV will provide high luminosity for high accuracy
measurements to test models on a large $x_{B}$ scale and thus will be a great
facility to study GPDs in the valence quark region. The new CLAS12
spectrometer, with its large acceptance, will be well suited for a rich and
exciting GPD experimental program.

\section*{Acknowledgments}
Thanks to P. Stoler and R. Ent for the opportunity to give this
presentation. Thanks to M. Guidal and S. Niccolai for useful discussions
and for providing slides used for the preparation of this talk.


\begin{thebibliography}{99}

\bibitem{gpd1}
D. M\"{u}ller, D. Robaschik, B. Geyer, F.-M. Dittes, and J. Horejsi, Fortschr. Phys. {\textbf 42}, 101 (1994).

\bibitem{gpd2}
X. Ji, Phys. Rev. Lett. 78, 610 (1997); Phys. Rev. D {\textbf 55}, 7114 (1997).

\bibitem{gpd3}
A.V. Radyushkin, Phys. Lett. B 380 (1996) 417; Phys. Rev. D {\textbf 56}, 5524 (1997).

\bibitem{gpd4}
J.C. Collins, L. Frankfurt and M. Strikman, Phys. Rev. D {\textbf 56}, 2982 (1997).

\bibitem{gpd5}
K. Goeke, M.V. Polyakov and M. Vanderhaeghen, Prog. Part. Nucl. Phys. {\textbf 47}, 401 (2001).

\bibitem{gpd6}
M. Diehl, Phys. Rept. {\textbf 388}, 41 (2003).

\bibitem{gpd7}
A.V. Belitsky, A.V. Radyushkin, Phys. Rept. {\textbf 418}, 1 (2005).

\bibitem{gpd8}
M. Guidal, Prog. Part. Nucl. Phys. {\textbf 61}, 89 (2008).

\bibitem{vgg1}
M. Vanderhaeghen, P.A.M. Guichon, and M. Guidal, Phys. Rev. D {\textbf 60}, 094017 (1999).

\bibitem{vgg2}
M. Guidal, M.V. Polyakov, A.V. Radyushkin and M. Vanderhaeghen, Phys. Rev. D {\textbf 72}, 054013 (2005).

\bibitem{bmk}
A. Belitsky, D. M\"{u}ller and A. Kirchner, Nucl. Phys. B {\textbf 629}, 323 (2002).

\bibitem{bsa1}
S. Stepanyan {\textit et al.} (CLAS Collaboration), Phys. Rev. Lett. {\textbf 87}, 182002 (2001).

\bibitem{bsa2}
G. Gavalian {\textit et al.} (CLAS Collaboration), Phys. Rev. C {\textbf 80}, 035206 (2009).

\bibitem{tsa}
S. Chen {\textit et al.} (CLAS Collaboration), Phys. Rev. Lett. {\textbf 97}, 072002 (2006).

\bibitem{clas}
B. Mecking {\textit et al.}, Nucl. Instrum. Meth. A {\textbf 503}, 513 (2003).

\bibitem{bsa3}
F.X. Girod {\textit et al.} (CLAS Collaboration), Phys. Rev. Lett. {\textbf 100}, 162002 (2008).

\bibitem{jml}
J.M. Laget, Phys. Rev. C {\textbf 76}, 052201(R) (2007).

\bibitem{hsj}
H.S. Jo, Ph.D. thesis, Universit\'{e} Paris-Sud, Orsay, France (2007).

\bibitem{guidal_clas}
M. Guidal, Phys. Lett. B {\textbf 689}, 156 (2010).

\bibitem{halla}
C. Munoz Camacho {\textit et al.} (JLab Hall A Collaboration), Phys. Rev. Lett. {\textbf 97},
262002 (2006).

\bibitem{guidal_fitter_code}
M. Guidal, Eur. Phys. J. A {\textbf 37}, 319 (2008) [Erratum-ibid. A {\textbf 40}, 119 (2009)].

\bibitem{moutarde}
H. Moutarde, Phys. Rev. D {\textbf 79}, 094021 (2009).

\bibitem{guidal_moutarde}
M. Guidal and H. Moutarde, Eur. Phys. J. A {\textbf 42}, 71 (2009).

\bibitem{guidal_hermes}
M. Guidal, Phys. Lett. B {\textbf 693}, 17 (2010).

\bibitem{km}
K. Kumeri{\v c}ki and D. M\"{u}ller, Nucl. Phys. B {\textbf 841}, 1 (2010).

\bibitem{km2}
K. Kumeri{\v c}ki and D. M\"{u}ller, arXiv:1008.2762 [hep-ph].

\bibitem{ps1}
R. De Masi {\textit et al.} (CLAS Collaboration), Phys. Rev. C {\textbf 77}, 042201(R) (2008).

\bibitem{ps2}
K. Park {\textit et al.} (CLAS Collaboration), Phys. Rev. C {\textbf 77}, 015208 (2008).

\bibitem{rho01}
C. Hadjidakis {\textit et al.} (CLAS Collaboration), Phys. Lett. B {\textbf 605}, 256-264 (2005).

\bibitem{rho02}
S. Morrow {\textit et al.} (CLAS Collaboration), Eur. Phys. J. A {\textbf 39}, 5-31 (2009).

\bibitem{omega}
L. Morand {\textit et al.} (CLAS Collaboration), Eur. Phys. J. A {\textbf 24}, 445-458 (2005).

\bibitem{phi1}
K. Lukashin {\textit et al.} (CLAS Collaboration), Phys. Rev. C {\textbf 63}, 065205 (2001).

\bibitem{phi2}
J. Santoro {\textit et al.} (CLAS Collaboration), Phys. Rev. C {\textbf 78}, 025210 (2008).

\bibitem{rho+}
A. Fradi, Ph.D. thesis, Universit\'{e} Paris-Sud, Orsay, France (2009).

\bibitem{gk}
S.V. Goloskokov and P. Kroll, Eur. Phys. J. C {\textbf 42}, 281 (2005); Eur. Phys. J. C {\textbf 50}, 829 (2007).

\bibitem{gk_phi}
S.V. Goloskokov, arXiv:0910.4308 [hep-ph].

\end{thebibliography}
\end{document}